# Meteorites and Planet Formation

**Rhian H. Jones**

*Department of Earth and Environmental Sciences*
*The University of Manchester*
*Manchester, M13 9PL, U.K.*
*Rhian.jones-2@manchester.ac.uk*

## ROLE OF METEORITES IN UNDERSTANDING PLANET FORMATION

Meteorites are a remarkable resource. They capture the imagination of people worldwide with their spectacular entry through Earth's atmosphere as fireballs, and their exotic character of being pieces of other worlds. Scientifically, they are critical to interpreting the early stages of formation of the Solar System, as well as the geological evolution of asteroids, the Moon, and Mars, and they are vital to understanding planetary formation processes. With the burgeoning exploration of extrasolar planetary systems, knowledge of the fundamental process of planetary growth from protoplanetary disks has taken on a new significance. Meteorites provide essential and detailed insight into the formation of planetary systems, although we must bear in mind that they only represent one reference point (our own Solar System) in what is clearly a wide spectrum of possible chemical and physical parameters governing the diverse realm of extrasolar planets. This chapter summarises the nature of our meteorite collections, and the ways in which meteorites contribute to our understanding of the formation and evolution of our own Solar System, with broader implications for planetary systems in general.

Within our own Solar System, each planet and minor body (such as the Moon, Pluto, Ceres) has a unique composition and history. The asteroid population is also extremely diverse in terms of chemistry and geologic history, within the main asteroid belt as well as in other populations such as Jupiter's trojan asteroids. There are many fundamental unknowns relating to these observations, and aspects of our own Solar System that are not well understood. For example: How were the different chemical and isotopic compositions of individual rocky planets established? How do current planet compositions compare with initial compositions shortly after planet formation? What is the volatile element inventory of planetary mantles? When and how was water incorporated into the terrestrial planets? What role did Jupiter play in controlling the compositions of inner and outer Solar System material? Meteorites play an important role in addressing such questions. This chapter mainly focuses on the meteorites that derive from asteroids, and their importance for understanding the early stages of formation of our planetary system. The mineralogy and isotope geochemistry of these meteorites provide a tangible record of physical and chemical environments in the protoplanetary disk, and the nature of potential planet-building materials. In addition to asteroidal meteorites, lunar and martian meteorites provide essential geochemical information about other planetary bodies, enabling us to compare the chemistry of two planets, Earth and Mars, as well as informing us about topics such as the importance of magma oceans in the early stages of planet formation, the effect of giant impacts on planetary evolution, and crustal and surface processes on airless bodies.

## METEORITES FROM ASTEROIDS

In order to discuss the relationship between meteorites and the protoplanetary disk environment, it is necessary to explain the classification of meteorites and the essential characteristics of the different meteorite groups. There are two important divisions of asteroid-derived meteorites: undifferentiated and differentiated meteorites. The term differentiated refers to the separation of a body into distinct chemical layers. In the case of meteorite parent bodies, differentiation is the result of heating that has produced extensive melting, allowing for density-driven separation of a metallic core and rocky mantle. In contrast, the parent body of an undifferentiated meteorite has not melted. As a result, undifferentiated meteorites, known as chondrites, preserve protoplanetary disk materials. An important heat source in meteorite parent bodies was decay of the short-lived isotope $^{26}$Al (half-life 0.717 million years), which was present in sufficient abundance in early-formed planetesimals to drive extensive melting (McSween et al. 2002). The potency of this energy source diminished towards the later stages of disk evolution, leading to milder heating of chondrite parent bodies (e.g. Desch et al. 2018). Impacts are an alternative heat source (e.g. Rubin 1995), but they are unlikely to account for the global-scale heating of asteroids recorded by meteorites (McSween et al. 2002). However, even though the collision rate has decreased throughout Solar System history, there is no a priori time limitation to the effects of impact-related heating.

**Undifferentiated meteorites: Chondrites**

Chondrites are the most abundant type of meteorites in our collections, 90% of all meteorites (which total around 73,000 to date: Meteoritical Bulletin Database). There are three major classes of chondrites: ordinary, carbonaceous and enstatite classes, and each class is divided into multiple groups and subgroups (Krot et al. 2014; Scott and Krot 2014). There are additional groups which do not fit into any of these classes (Kakangari-like (K) chondrites, Rumuruti-like (R) chondrites), as well as individual anomalous chondrites, but we will not concern ourselves with the details here. For the purposes of this discussion, it is important to understand divisions at the level of the major classes and their comprised groups. The essential characteristics of these groups are described below, following a brief summary of the basic chondrite components.

Chondrites are made up of a set of components which have very different compositions and origins (e.g. Scott and Krot 2014). Chondrites are named after their most abundant component, chondrules (Fig. 1). These are (sub-) millimeter-sized beads that formed from individual melt droplets (Connolly and Jones 2016; Jones 2012). Chondrules typically consist of the (iron-magnesium-calcium) silicate minerals olivine and pyroxene, as well as either glassy or crystalline interstitial "mesostasis" material that has a feldspathic composition (feldspar is a calcium-sodium aluminosilicate mineral). Many chondrules also include beads of iron-nickel metal (Fe,Ni) and iron sulfide (FeS). Individual chondrules have a wide range of chemical compositions and oxidation states, with the Mg/(Mg+Fe) atomic ratio in olivine and pyroxene varying from 0.99 to 0.70 or lower. The heating mechanism that produced chondrule melt droplets is a matter of debate. One of the most favoured current models proposes heating of dust in shock fronts, such as in the bow shock of planetesimals as they migrated through the disk (Morris and Boley 2018). Planetesimal collision models are also considered (Johnson et al. 2018). Heating models for chondrule formation are necessitated by the fact that chondrules exist: if we did not observe them in chondrites, they would not be predicted in our own or any other planetary system, based on current observations and disk formation models.

The second dominant component of chondrites is matrix (Fig. 1). This is a fine-grained, sub-micrometer, mixture of different materials which includes low-temperature components that must have escaped chondrule heating events (Scott and Krot 2014). Matrix surrounds and supports chondrules, and also occurs as well-defined fine-grained rims adhering to chondrules. In chondrites that represent the most primitive materials, i.e. those that have not been affected by the secondary processes discussed below, matrix consists predominantly of crystalline and amorphous silicates, with abundant nano-scale sulfides and carbides (Abreu and Brearley 2010; Dobrică and Brearley 2020). Because of its fine-grained nature, matrix is highly susceptible to alteration, and this primitive mineralogy is only retained in a few rare chondrites. Carbon-rich materials are also an important matrix component, including organic material that consists of insoluble macromolecular carbon and soluble organics such as amino acids (Grady and Wright 2003). Matrix is also the host of presolar grains, which are identified by their non-solar isotopic compositions (e.g. noble gases, Si, Ti, C, O, N isotopes), and which include graphite, diamond, silicon carbide, oxides, and presolar silicates (Nittler and Ciesla 2016). Primitive matrix therefore is an accumulation of dust particles that represents a low-temperature protoplanetary disk environment. Because we are unable to recognise presolar material with solar isotopic compositions, the proportion of matrix material that is inherited directly from the presolar molecular cloud is unknown.

Refractory inclusions are a less abundant (up to 3 vol%, but more typically less than 1 vol% in most chondrite groups: Scott and Krot 2014), but nevertheless highly significant, chondrite component (MacPherson et al. 2014; Krot 2019: Figs. 1a,d). Two main types of refractory inclusions are calcium-, aluminum-rich inclusions (CAIs) and amoeboid olivine aggregates (AOAs). CAIs consist of highly refractory minerals, in the sense that these minerals are the first to condense (and last to vaporize) from a gas of solar composition. Their constituent minerals, rich in Ca, Al, Si, Ti and O include perovskite, hibonite, spinel, melilite, pyroxene, anorthite and a multitude of other minerals that are highly exotic to a planetary geologist. In addition, CAIs have distinct oxygen isotope compositions, which range to $^{16}$O-rich values that approach the oxygen isotope composition of the Sun (see below). The oldest absolute ages measured in Solar System material are Pb-Pb ages of CAIs, 4.567 Ga (billion years) (Connelly et al. 2012). Thus, CAIs are thought to represent the first solids that were produced within the Solar System. CAIs are formed in hot, highly reducing environments, possibly at the inner edge of the protoplanetary disk, although their subsequent distribution throughout the disk must be accounted for in disk models, as well as the timing of formation relative to chondrules, and the relationship of CAIs to the two isotopic reservoirs of Solar System materials discussed below (Cuzzi et al. 2003; Desch et al. 2018). CAI-like inclusions, as well as chondrule-like inclusions, have been identified in samples from comet Wild 2, which implies a disk-wide transport mechanism (Joswiak et al. 2017).

In order to understand the distinctions between chondrite groups it is important to understand the framework of alteration that individual meteorites have undergone. This is described in a scheme of "petrologic types". The most primitive chondrites, i.e. those that have undergone little alteration since accretion of their parent bodies, are classified as petrologic type 3, and the least altered of these are subtype 3.00, on a scale of 3.00 to 3.9. Progressive effects of parent-body heating are denoted as petrologic types 3, 4, 5 and 6 (and sometimes 7). Progressive effects of aqueous alteration on chondrite parent bodies are described on a scale from petrologic type 3 (unaltered) to 2 then 1 (completely altered).

Ordinary chondrites (OCs) (Fig. 1b,c) are the most abundant type of meteorites to fall (94% of all chondrites). They have a high chondrule / matrix ratio, typically consisting of around 80% chondrules. There are three main OC groups, H, L and LL, which have high (H) or low (L) amounts of metallic iron, or low metallic iron as well as low total iron (LL). Abundances of metallic iron vary from 8 to 3 to 1.5 vol% in H, L and LL chondrites respectively (Scott and Krot 2014), which reflects differences in bulk chemical composition as well as oxidation state on their parent asteroids. It is likely that the H, L, and LL chondrites are derived from separate parent asteroids (Kallemeyn et al., 1989; Krot et al. 2014). Most ordinary chondrites were heated on their parent asteroids, in the first few million years after accretion (McSween et al. 2002). This resulted in thermal metamorphism (petrologic types 3 to 6), characterised by recrystallisation of the initial chondrule / matrix texture, chemical equilibration of minerals such as olivine and pyroxene that were initially very heterogeneous in composition, and formation of new minerals including feldspar, oxides, and phosphates (Huss et al. 2006). The early stages of heating occurred in the presence of a fluid (a process known as metasomatism), which included formation of water-bearing minerals such as phyllosilicates, and mobilisation of water-soluble elements (e.g. Lewis et al. 2022). Bulk water contents of OCs are typically less than 1 wt%, but it is important to recognise that most OCs have lost their initial water during heating (Fig. 2). Average water and carbon contents of the most primitive OCs (petrologic subtypes 3.00 to 3.2) are both around 0.5 wt% (Fig. 2). Peak metamorphic temperatures in OC parent bodies were around 950 °C, the temperature at which the Fe-Ni-S system begins to melt (the eutectic temperature). However, by definition a chondrite has not experienced any melting, so this is a somewhat self-fulfilling limitation. The extent to which the OC parent bodies might have melted in their interiors, producing material that we would not classify as chondritic, is unresolved. Paleomagnetic records show that some partially molten planetesimals might have had a core and mantle, overlain by a chondritic crust (e.g. Dodds et al. 2021).

Carbonaceous chondrites (CCs) (Fig. 1a,d) comprise 5% of all chondrites. They have lower chondrule / matrix ratios than OCs, typically consisting of around 50% chondrules. CC groups are named after a typical meteorite in that group, preferably an observed fall, for example the CI group is named after the Ivuna meteorite which fell in Tanzania in 1938. Several CC groups, notably the CI, CM and CR groups, have been altered, some extensively, by the action of water (aqueous alteration: petrologic types 1 and 2) as demonstrated by high abundances of phyllosilicates, carbonate minerals, and the iron oxide, magnetite (Fig. 1d). The most likely model is that water ice accreted initially to CC parent bodies, and then mild heating to <200 °C melted the ice, resulting in fluid flow (e.g. Vacher and Fujiya 2020). In heavily altered CCs, chondrules have been erased by alteration reactions, resulting in high apparent matrix abundances that are not necessarily reflective of the primary abundance. For example, CI chondrites (petrologic type 1) consist of 100% matrix, but they might have initially contained chondrules. It is important to recognise that not all CCs are aqueously altered in a similar way. CO, CV and CK chondrites have considerably lower abundances of phyllosilicates and carbonate minerals than CI, CM and CR groups. The CO and CV groups, mostly petrologic type 3, experienced higher-temperature metasomatism (up to 300 °C), and the CK group shows a range of metamorphic effects comparable to those observed in OCs (i.e. peak temperatures up to 950 °C, petrologic types 3 to 6). The CB and CH groups are distinct from other CCs, in that they are very metal-rich and their chondrules are younger than other chondrite groups. They are thought to be the products of a giant impact plume resulting from a collision between planetesimals (e.g. Krot et al. 2005, 2014).

Overall, the carbonaceous chondrite class is very diverse (Braukmüller et al. 2018). The bulk composition of the CI group is a very close match to the solar photosphere (Lodders 2003). However, other CC groups are relatively depleted in moderately volatile elements such as Mn, Na and K, for example the bulk Mn/Mg ratio in CM chondrites is around 0.7 x the ratio in CI chondrite (Lodders and Fegley 20; Scott and Krot 2014). Measured bulk water contents of CCs also vary widely, from <1 to more than 10 wt% $H_2O$, in aqueously altered CI, CM and CR chondrites (Fig. 2). Calculations of accreted water abundances for carbonaceous chondrites range between 3% to 28% $H_2O$ for CO and CI groups respectively (Alexander 2019). The wide range of measured values is attributable to different degrees of secondary alteration, and whether a chondrite retains the products of aqueous alteration reactions. For example, CM chondrites record a range of degrees of alteration, from essentially unaltered to almost completely altered. In addition, a number of CM chondrites, as well as the CY chondrites, have undergone heating that resulted in dehydration following aqueous alteration, resulting in lower $H_2O$ abundances (King et al. 2019, 2021). Terrestrial contamination is a potential problem when $H_2O$ abundances are measured (Vacher et al. 2020). The $H_2O$ content of pristine CI-like material returned from asteroid Ryugu, 6.8 wt%, is much less than that measured in the same study in the Ivuna CI chondrite (12.7 wt%). This difference is interpreted as terrestrial adsorption of $H_2O$ in CI chondrites, most of which are old falls (Yokoyama et al. 2023). $H_2O$ contents of CI and CM chondrites determined by Vacher et al. (2020) and Piani et al. (2021), in which adsorbed water is degassed prior to analysis, are mostly less than 10 wt%.

Since carbon is initially concentrated in matrix, and CCs have high matrix abundances, many CCs have high bulk carbon contents as their name implies. Carbon was introduced into chondrite parent bodies as organic material and / or $CO/CO_2$ ices (Alexander et al. 2012, 2015). CI chondrites and the ungrouped, highly altered carbonaceous chondrite, Tagish Lake, have C contents up to 4 wt% (Fig. 2), and CI-like material from asteroid Ryugu has 4.6 wt% C (Yokoyama et al. 2023). However, not all carbonaceous chondrites have high bulk carbon content, for example carbon contents of thermally altered CO and CV groups overlap with carbon contents of OCs, around 0.5 wt% C (Fig. 2).

Enstatite chondrites (ECs) are a rarer class of chondrites, 1% of all chondrites. Their chondrule / matrix ratios are comparable to OCs, up to 80% chondrules. The EH and EL groups have high and low amounts of metal (plus sulfide), respectively, and high abundances of the Mg-rich pyroxene, enstatite (hence their name). They are rich in sulfur (3-6 wt% in bulks: Lodders and Fegley) and EH chondrites in particular have higher abundances of moderately volatile elements such as Mn, Na, K, and Cl than most carbonaceous chondrite groups (Scott and Krot 2014; Brearley and Jones 2018). The ECs are extremely reduced rocks: the Mg/(Fe+Mg) atomic ratio of olivine and pyroxene is very high, >0.99. They also contain an exceptionally diverse array of sulfide minerals, including Mn, Ca and K sulfides that are otherwise very rare in planetary environments because these are lithophile (rock-seeking) elements that are normally found in silicate and oxide minerals. Carbon contents of ECs are 0.2 – 0.7 wt%, and much of the carbon is present as graphite (Grady and Wright 2003; Alexander et al. 2007; Storz et al. 2021). Measured hydrogen abundances of ECs give equivalent water contents of <1 wt% (Piani et al. 2020), but it is not clear that H is present as $H_2O$. The ECs show the same range of metamorphism as OCs (petrologic types 3 to 6), which resulted in chemical and textural equilibration to the point where the initial chondrule / matrix texture was erased. Because they are highly reduced, ECs are thought to originate in the inner regions of the protoplanetary disk where little $H_2O$ was available to oxidise the environment.

Chondritic meteorites are derived from asteroids, and matches between meteorites and asteroid types can be made from reflectance spectra. In general, OCs are derived from S-type, CCs from C-type, and ECs from Xc-type spectral classes (DeMeo et al. 2022), but the details are quite complex. Some meteorite-asteroid connections have been confirmed by recent return missions of asteroid samples: the JAXA Hayabusa mission to S-type asteroid Itokawa returned samples similar to LL ordinary chondrites (Nakamura et al. 2011), and the JAXA Hayabusa2 mission to Cb-type asteroid Ryugu returned samples similar to CI carbonaceous chondrites (Yokoyama et al. 2022). The results of samples returned by the NASA OSIRIS-REx mission to B-type asteroid Bennu in September 2023 will be known soon. The asteroid population in the main asteroid belt is zoned, with S-type asteroids concentrated towards the inner part of the belt and C-types concentrated at greater distances (DeMeo et al. 2014). The presence in the main asteroid belt of chondritic asteroid bodies with very different compositions, matrix abundances, and oxidation states, which are thought to have originated across a wide span of heliocentric distances, is thought to be attributable to perturbance of the initial planetesimal distribution, caused by migration of the giant planets (Walsh 2011; Morbidelli et al. 2015). An important aspect of such a model is that it could have brought material that formed in the outer regions of the disk into the inner Solar System, a point that is discussed further below. It is important to recognise that the distribution of meteorites in our collections is not representative of the asteroid population, which is dominated by CC material. Several factors contribute to this sampling bias, including recent breakup of asteroids, proximity of asteroids to the Kirkwood gap orbital resonances that enhance delivery to Earth-crossing orbits, and the susceptibility of friable materials to fragmentation during asteroid collisions and atmospheric entry.

**Differentiated meteorites: Achondrites and irons**

The term differentiated meteorites encompasses a wide range of meteorite types, all sourced from parent bodies (planetesimals) on which melting and core formation processes have taken place. At a basic level, the main types are iron meteorites, derived from planetesimal cores, and achondrite meteorites, which are rocky (stony) materials derived from mantles and crusts. (The word achondrite literally indicates simply that a meteorite is not a chondrite, but it is commonly used to refer to only the rocky material of a differentiated planetesimal.) There are also various groups of "primitive achondrites", which derive from bodies that have seen less extensive melting, and which we will not discuss here. An additional division of differentiated meteorites, named stony-irons, consists of meteorites that contain approximately equal abundances of stony material (i.e. silicate minerals) and Fe,Ni metal. This includes the pallasite and mesosiderite groups. However, the stony-iron descriptor includes materials with very different origins, and does not uniquely identify a formation process. Pallasites consist of approximately equal abundances of iron metal and olivine. Various origins have been proposed for pallasites, including mixing at the core / mantle boundary of a differentiated planetesimal, disruption of a partly solidified planetesimal, and ferrovolcanism arising from a partially molten core (Yang et al. 2010; Boesenberg et al. 2012; Johnson et al. 2020). Disruptive impacts have played a prominent role in the evolution of differentiated bodies: several groups of irons (classed as non-magmatic irons), as well as mesosiderites, are the products of significant mixing of material from different depths in their parent bodies, and their impactors, during the period when parent bodies were partially molten (Goldstein et al. 2009; Scott et al. 2001; Haba et al. 2019). Iron meteorites that represent the relatively undisturbed products of core formation, named magmatic irons, have of course subsequently been excavated from the depths at which they originally formed, by collisional erosion over an extended time period. The following discussion focuses on achondrites and irons, which are the essential groups needed to further our understanding of planet-formation processes in the early Solar System.

**Achondrites.** Achondrites are igneous rocks that crystallised from silicate melts, either at the surface (volcanic or extrusive rocks) or at depth (intrusive rocks) in their parent planetesimals. Silicate melts originate from partial melting of a solidified mantle, and may undergo processes such as fractionation (i.e. partial crystallisation, with removal of the solid fraction) or assimilation (i.e. incorporation of surrounding rock) as they rise to the surface of the body. In addition to the range of chemical compositions that can be attributed to such processes, achondrites in our meteorite collections have diverse compositions that attest to a wide range of mantle compositions and oxidation states. These chemical groupings, coupled with oxygen isotope compositions (see below), define different achondrite groups that likely each originate from a different parent body. The origin of this primary diversity among differentiated planetesimals is likely related to location and timing of accretion within the protoplanetary disk. The requirement for localised accretion of diverse planetesimals that all currently reside within the main asteroid belt must be taken into account in disk models. Here we will focus on a few selected achondrite groups to illustrate the diversity, avoiding considerable complexity and detail that can be found elsewhere.

The best understood achondrites are the HED suite (e.g. Mittlefehldt 2014). The acronym stands for howardite, eucrite and diogenite, three achondrite groups that are closely related chemically, petrologically, and isotopically. They are the most abundant achondrites (75% of all achondrites, out of a total of around 3,800 known achondrites). Eucrites are basaltic rocks (Fig. 3a,b), and the group includes volcanic rocks that erupted as surface flows as well as intrusive cumulates. The dominant silicate minerals are pyroxene and plagioclase feldspar. Diogenites consist predominantly of pyroxene, specifically low-calcium orthopyroxene, and they represent cumulates that crystallised at depth. Many eucrites and diogenites are highly brecciated (i.e. they consist of an aggregate of broken rock fragments), as a result of extended impact bombardment at the surface of their parent asteroid (Fig. 3b). The howardites are mechanical mixtures of eucrite and diogenite material, including rock clasts and individual mineral grains, that were formed as surface breccias. The various compositions of the HED achondrites are consistent with derivation from the V-type asteroid 4 Vesta, which was the subject of intense remote study by NASA's Dawn mission (McCoy et al. 2015). This connection is the strongest link between a group of achondrites and a known asteroid source.

The ureilites are also an abundant group of meteorites (18% of achondrites). These achondrites are composed predominantly of pyroxene and olivine, and they represent restites (the residues after melts have been extracted) and cumulates derived from a carbon-rich parent body (e.g. Mittlefehldt et al. 2014; Goodrich et al. 2009). They contain carbon in the form of graphite and diamond, and they have diverse oxygen isotope ratios related to carbonaceous chondrites (see below).

The angrites are a rather rare group of achondrites (1% of achondrites). They are basaltic, predominantly consisting of pyroxene, olivine, and plagioclase feldspar (Fig. 3c), and include volcanic as well as intrusive (cumulate) rocks; vesicles (gas bubbles) are common in volcanic angrites (e.g. Mittlefehldt et al. 2014). Angrites are enriched in refractory elements (i.e. elements with high condensation temperatures) and have unusual Al,Ti-rich pyroxene and Ca-rich olivine compositions.

They are extremely depleted in moderately volatile elements such as Na and K, which appears to be controlled by partial condensation of materials accreted to the angrite parent body (Tissot et al. 2022). Unlike most other achondrites, and notably the eucrites, many angrites are remarkably unshocked and unbrecciated: this could be attributed to early catastrophic impact and breakup of the original parent body (Scott and Bottke 2011).

The enstatite achondrites (2% of achondrites), also known as aubrites, are the most highly reduced igneous rocks in the Solar System. Like their chondrite counterparts, they consist predominantly of enstatite, a pyroxene mineral with a very high Mg/(Mg+Fe) atomic ratio >0.99 (e.g. Mittlefehldt et al. 2014). Aubrites and enstatite chondrites have similar oxygen isotope ratios (see below) and similar chemical compositions, but several aspects of these rocks rule out formation of aubrites directly from an EC parent body (Keil 2010). Most aubrites are highly brecciated.

**Iron meteorites.** As mentioned above, iron meteorites are described as either magmatic or non-magmatic irons (e.g. Goldstein et al. 2009). Iron meteorites consist predominantly of iron-nickel metal (Fe,Ni) that occurs as two main minerals, kamacite (Ni <5 wt%) and taenite (Ni >10 wt%). Iron sulfide, FeS (mineral name troilite), is also common, as are minor phosphides, carbides, and graphite. Iron meteorites are classified according to two different criteria. One criterion is based on the (micro)structure of the Widmanstätten pattern, an intergrowth of kamacite and taenite that is produced during slow cooling (over millions of years) in the solid state, from solid metal of uniform composition (Fig. 3d). This gives rise to classifications such as octahedrite and hexahedrite. Cooling rates of iron meteorites, determined from stranded chemical diffusion profiles in kamacite and taenite, indicate that their parent bodies were hundreds of kilometers in diameter (Goldstein et al. 2009; Benedix et al. 2014). The second criterion is a chemical classification, based on the bulk composition of the sample. Discriminatory classification diagrams of Ir vs Ni and Ge vs Ni content (Fig. 5) are used to identify different chemical groups that have names such as IAB. There are also numerous ungrouped iron meteorites that do not fit into well-defined groups. Based on these classifications, it appears that the (currently) 1,375 known iron meteorites represent the cores of numerous differentiated planetesimals: more than 50 distinct bodies are likely sampled (Benedix et al. 2014). The asteroid parent bodies of iron meteorites are thought to be the M-type asteroids, one of which, 16 Psyche, is the target of NASA's current Psyche mission (Elkins-Tanton et al. 2022). Since reflectance spectra of metals are very featureless, and all iron meteorites consist of the same basic mineralogy, there is little possibility of matching the different iron groups to individual asteroids.

Examples of magmatic irons include the IIAB, IIC, IIIAB, IVA and IVB groups. Within each magmatic iron group, the chemical compositions of individual meteorites show a well-defined, negative and often steep trend on the Ir vs Ni classification diagram (Fig. 5a), consistent with fractional crystallisation of solid metal from liquid metal as the planetesimal cools and the core solidifies. Each magmatic iron group defines a different trend, with a range of Ni abundances from about 5 to 20 wt%, indicating different initial chemical compositions of the individual planetesimal cores. The Ge vs Ni classification diagram (Fig. 5b) illustrates differences in volatile element compositions, since Ge is a volatile element (50% condensation temperature, $T_c$, is 883 K: Lodders 2003). Ge contents of the magmatic irons vary over four orders of magnitude, which is attributed to significant differences in the bulk compositions of differentiated planetesimals.

The non-magmatic irons include the IAB complex and the IIE group: together these groups represent around 30% of all iron meteorites. These meteorites have high abundances of iron-nickel metal, but they are brecciated and contain silicate clasts. As mentioned above, disruptive impacts played a significant role in forming these rocks. Non-magmatic irons show compositional clusters on the Ir vs Ni and Ge vs Ni classification diagrams that are not related to solidification processes. Ir contents are in the range 1-10 ppm, but Ge contents are more variable. The IAB complex shows a wide range of Ge vs Ni, and silicate compositions are broadly chondritic, suggesting formation on a partially differentiated parent body (Wasson and Kallemeyn 2002; Benedix et al. 2014).

## PLANETARY METEORITES FROM THE MOON AND MARS

### Lunar meteorites

There are over 600 known lunar meteorites, although many of these are paired and the number of unique meteorite falls is much lower. Lunar meteorites include mare basalts, highland anorthosites, breccias that contain multiple lithologies, and impact melt breccias (e.g. Korotev 2005). These correspond to the most abundant lunar rock types known from remote measurements of the lunar surface and from returned samples. They are typically highly shocked, as a result of the impacts that lofted them from the Moon's surface: for example, feldspar is commonly transformed to the amorphous shocked phase maskelynite in lunar meteorites, indicating shock pressures 20-30 GPa (Rubin 2015; Chen et al. 2019: Fig. 3e). Lunar meteorites are very important to our understanding of lunar evolution (Korotev 2005; Joy and Arai 2013; Tartèse et al. 2019). In contrast to returned samples from the Apollo, Luna and recent Chang'e 5 missions, their original locations on the Moon's surface are unknown. However, the random nature of their source locations means that they provide a more accurate overall picture of the Moon's bulk chemistry, very likely including high-latitude regions and the far side of the Moon that have not been sampled by missions to date. Lunar meteorites play an important role in interpreting the impact history of the Moon, including discussion of the "late heavy bombardment", a proposed period of intense impact events at around 3.9 Ga that would have affected all of the terrestrial planets, not only the Moon, and which has implications for the early development of life (e.g. Bottke and Norman 2017).

### Martian meteorites

There are around 360 known martian meteorites, although as for lunar meteorites, many of these are paired and the number of unique meteorite falls is lower. Most martian meteorites belong to a suite of geochemically related mafic and ultramafic igneous rocks, including volcanic basalts, and intrusive gabbros, olivine cumulates, and pyroxene cumulates (e.g. McSween 2015: Fig. 3f). Their relatively young geologic ages (as young as 200 Ma), similarity to rocks on the martian surface observed by remote sensing and lander missions, and the match of their gas compositions to the martian atmosphere, provide convincing evidence that they originate from Mars (McSween 2015). Like lunar meteorites, most martian meteorites were highly shocked during the impact that lifted them from Mars, including transformation of plagioclase to maskelynite. More recently, an important group of martian regolith breccias has been recognised that probably originate from the ancient martian highland region (Agee et al. 2013; Cassata et al. 2018; McCubbin et al. 2019). These rocks contain clasts (fragments) of numerous rock types, including more geochemically evolved rocks than the main groups of martian meteorites which expand the available geological record significantly. Martian meteorites contain important evidence for volatile element abundances and hydrothermal activity on Mars (e.g. McCubbin and Jones 2015; Filiberto et al. 2016).

Since martian meteorites are currently our only samples of Mars, they are a unique resource for investigating the isotopic composition, mantle chemistry, geologic evolution, volatile content, and surface processes of a different planet in the Solar System, giving us two data points rather than just one (the Earth) for a detailed comparison of the geochemical properties of planets. The details of these topics are mostly beyond the scope of this article. Isotopic compositions of martian meteorites are included in the discussion below.

## CHEMICAL AND ISOTOPIC HETEROGENEITY IN THE SOLAR SYSTEM

### Chemical heterogeneity in the inner Solar System

From the above discussion, it is apparent that there is considerable chemical variability in the compositions of rocky bodies in the inner Solar System, including planets, the Moon, and asteroids. A detailed discussion of the bulk chemistry of individual meteorite parent bodies and planets is beyond the scope of this review (see Putirka 2024, this volume). Some of the complexity in understanding the chemical differences among the mantles of differentiated bodies can be illustrated with the following example. Basaltic rocks from different meteorite parent bodies, including the planetary bodies Earth, Moon and Mars, have clearly defined Mn/Fe ratios in their silicate minerals, olivine and pyroxene (Fig. 5). The different Mn/Fe ratios are well enough constrained that they are a useful parameter to classify the parentage of basaltic meteorites. Since Mn and Fe behave similarly in igneous processes, differences in Mn/Fe must reflect distinctly different bulk Mn/Fe ratios in planetesimal or planetary mantles. Such differences may partly be attributable to the fundamental chemistry of the body: since Mn is more volatile than Fe (50% condensation temperatures, $T_c$, for Mn and Fe are 1158 K and 1334 K respectively: Lodders 2003), the higher Mn/Fe ratio of Mars than Earth could be attributed to accretion at a greater radial distance from the Sun, where disk temperatures were lower, and more Mn had condensed (e.g. Papike et al. 2003). Similarly, the lower Mn/Fe ratio of the Moon is consistent with loss of volatiles during the Moon-forming giant impact. If volatility is the main control on Mn/Fe ratios, asteroid 4 Vesta would have formed closer to Mars, and the angrite parent body closer to Earth (Fig. 5: Papike et al. 2017). However, an additional factor is the oxidation state of the planet: a higher degree of reduction will lead to greater sequestering of metallic Fe into a planetary core which will raise the initial Mn/Fe ratio of the derivative mantle (Papike et al. 2017). As a result, we would expect a higher Mn/Fe ratio closer to the Sun which counteracts the volatility trend. Although it is not clear which factor exerts more control, the observation that each body has a unique mantle Mn/Fe ratio clearly points to chemical heterogeneity among planetary bodies.

One of the key questions related to the composition of a planet is the abundance of water, as well as other volatiles, including carbon, nitrogen, and the halogen elements, fluorine and chlorine (e.g. Broadley et al. 2022; Halliday and Canup 2023; Lodders and Fegley 2023). Water plays a major role in mantle and crust evolution, and it is of course essential for life. Water is also critical in determining the oxidation state of a planet, which in turn determines the relative size of mantle and core, and mantle chemistry. As discussed above, most primitive chondritic materials, including ordinary as well as carbonaceous chondrites, contain evidence for initial presence of water, which probably accreted to planetesimals as water ice. In addition, achondrite parent bodies, including the eucrite, angrite and ureilite parent bodies, initially contained water at abundances of tens to hundreds of ppm (e.g. McCubbin and Barnes 2019; Peterson et al. 2023). Isotopic compositions of hydrogen (D/H ratios), as well as C and N isotopes, contribute to debate about whether achondrite parent bodies accreted water during planetesimal growth, or whether water was added later in the form of carbonaceous chondrite material, as well as the extent of planetary degassing and the contribution of interstellar water (e.g. McCubbin and Barnes 2019; Newcombe et al. 2023). Similarly, the source of water on the Earth and other terrestrial planets has been proposed to be either from late accretion of a cometary or carbonaceous chondrite-like component, or from primary accretion of inner Solar System material (Alexander et al. 2012, 2018; Marty 2012; Piani et al. 2020; Broadley et al. 2022; Tissot et al. 2022; Izidoro and Piani 2022; Halliday and Canup 2023). Although these discussions continue to evolve, overall it is clear that water ice was a ubiquitous primary component of most of the planetary materials from rocky parent bodies that have been sampled to date, and that water and other volatiles can be considered as inevitable components of planets, at least in our own planetary system.

**Oxygen isotope compositions of planetary materials**

Oxygen isotope compositions of meteorites show significant diversity, and they are an essential aspect of understanding the evolution of the protoplanetary disk and our planetary system (Ireland et al. 2020). Oxygen isotope compositions are typically illustrated on an oxygen three-isotope diagram (Fig. 6a-c), with compositions given in the delta notation such that $^{17}O/^{16}O$ and $^{18}O/^{16}O$ ratios are referenced to a terrestrial standard, standard mean ocean water (SMOW) by the relationship: $\delta^{17,18}O = [(^{17,18}O/^{16}O)_{sample} / ^{17,18}O/^{16}O_{SMOW}) - 1] \times 1000$ ‰. On this diagram, the terrestrial fractionation (TF) line, which defines the oxygen isotope compositions of Earth materials, is used as a reference line. The TF line has a slope of 0.52, derived from geochemical reactions and processes that are controlled by mass fractionation effects. A useful construct is the definition of $\Delta^{17}O$, given as $\Delta^{17}O = \delta^{17}O - (0.52 \times \delta^{18}O)$, which defines the vertical offset of either a point or a mass-dependent fractionation line from the TFL on this diagram, as illustrated in Figure 6a. Two other reference lines are shown in Figures 6a-c: the carbonaceous chondrite anhydrous minerals (CCAM) line, which has a slope of 0.94, and the primitive chondrule mixing (PCM) line which has a slope of one, both of which are mass-independent fractionation or mixing lines (Dauphas and Schauble 2016), and which are discussed further below.

Chondrites show a range of oxygen isotope compositions among the different classes and groups (Fig. 6a). Bulk ordinary, enstatite and carbonaceous chondrites have bulk oxygen isotope compositions with decreasing values of $\Delta^{17}O$: enstatite chondrite values are similar to terrestrial compositions and lie on the TF line. For most chondrite groups, arrays of chondrite bulk compositions have slopes around one, parallel to the CCAM / PCM lines Within carbonaceous chondrite groups that have been affected extensively by aqueous alteration (CI, CM and CY groups), oxygen isotope compositions lie along arrays with shallower slopes as a result of reactions between primitive chondrule and matrix materials, and water. The bulk oxygen isotope composition of asteroid Ryugu, sampled by the Hayabusa2 mission, is similar to CI chondrites (Greenwood 2023; Tang et al. 2023).

Individual chondrules and refractory inclusions show remarkable heterogeneity in the oxygen 3-isotope diagram (Fig. 6b). Most bulk analyses of chondrules have similar compositional ranges to their respective bulk chondrite compositions. However, bulk analyses of refractory inclusions (CAIs and AOAs) lie on an array with a slope close to one: this observation originally defined the CCAM line (Clayton and Mayeda, 1999). The $^{16}O$-rich end-member of this array has $\delta^{17,18}O$ values around -45 ‰ (Fig. 6b). Individual in situ point analyses of mineral grains in refractory inclusions, made using secondary ion mass spectrometry (SIMS), also spread along the entire length of the CCAM line. For chondrules, in situ SIMS analyses of constituent mineral grains lie on a slightly different slope-1 array, the PCM line (Ushikubo et al. 2012). These observations are true for chondrules and refractory inclusions in ordinary as well as carbonaceous chondrites (e.g. Ushikubo et al. 2012; Williams et al. 2020; Piralla et al. 2021). Interpretation of these oxygen isotope distributions in primitive chondrite components is currently a matter of debate (Dauphas and Schauble 2016; Ireland et al. 2020). The $^{16}O$-rich end of the line is well understood, because it is similar to and consistent with the oxygen isotope composition of the Sun (McKeegan et al. 2011: Fig. 6b). However, the $^{16}O$-poor endmember is less well defined. The most accepted interpretation of the slope-1 distribution is that solar system solids exchanged oxygen with a "heavy", $^{16}O$-poor, component. This isotopic component could have been introduced into the disk in the form of $H_2O$ ice that originated from a region where dissociation of CO gas produced $H_2O$ ice enhanced in $^{17}O$ and $^{18}O$, as a result of CO self-shielding (e.g. Dauphas and Schauble 2016; Ireland et al. 2020). A possible identification of the $^{16}O$-poor component is a "cosmic symplectite" (COS: an intergrowth of the iron oxide, magnetite and the iron-sulphide mineral, pentlandite) observed in a primitive chondrite, which has $\delta^{18}O$ values of +180‰, lying on an extension of the slope-one lines (Sakamoto et al. 2007; Seto et al. 2008: Fig. 6b). However, this is a unique and rare occurrence. Further evidence for a similar $^{16}O$-poor component comes from hydrated interplanetary dust particles (IDPs) which likely record aqueous alteration in Kuiper belt bodies (Snead et al. 2017; Keller and Flynn 2022).

Among the achondrites, ureilites show considerable heterogeneity in their oxygen isotope compositions (Fig. 6c), which indicates that the ureilite parent body did not undergo global-scale melting (e.g. Greenwood et al. 2017). Apart from ureilites, most achondrites, as well as mesosiderites and (main group) pallasites, show well-defined mass-dependent fractionation trends along arrays of slope 0.52, parallel to the TF line, with separate fractionation lines for each group (Fig. 6c,d). For example, the HEDs, angrites, and aubrites have mean $\Delta^{17}O$ values of -0.24, -0.07 and +0.03 ‰ respectively (Greenwood et al. 2017). (Oxygen isotope compositions of mesosiderites overlap with those of the HEDs, and they may be derived from the same parent body, Vesta: Haba et al. 2019.) Martian meteorites also define a unique fractionation line, with a $\Delta^{17}O$ value of +0.32 ‰ (Franchi et al., 1999; Fig. 6c,d). The Moon has a near-identical $\Delta^{17}O$ value to the Earth (Hallis et al. 2010; Cano et al. 2020; Fig. 6c,d), which can be attributable to either similar oxygen isotope compositions of the Earth and the impactor, Theia, or mixing during the Moon-forming giant impact (Cano et al. 2020). These observations illustrate that Earth, Mars, possibly Theia, and the large asteroid 4 Vesta (thought to be the source of HEDs) each has a unique bulk oxygen isotope composition, and that oxygen isotope heterogeneity also occurs at a scale of individual differentiated planetesimals.

Oxygen isotope compositions of essentially all solid bodies that we have currently sampled lie within a limited range, with $\delta^{18}O$ values of most bulk meteorites lying between -5 and +10 ‰ (Fig. 6). This suggests an overall similar degree of exchange between solar values and the proposed $^{16}O$-poor component. It also requires that a large mass of material is affected by the exchange process, which in turn requires a large mass of isotopically heavy oxygen. An alternative explanation for the commonality of oxygen isotope compositions of asteroids and the terrestrial planets is that their composition represents the average dust from the molecular cloud, and that a highly $^{16}O$-poor water-ice component provides only a minor contribution (Ireland et al. 2020). Whatever the end-members of isotope exchange, refractory inclusions and chondrules capture the processes and environment in which exchange was taking place. The process that controls planet-scale and asteroid-scale heterogeneity in oxygen isotopes must relate to the timing and location of formation of primary accretionary building blocks, and the spatial extent of the accretionary zone from which each body is constructed. Although there is general acceptance of a model for oxygen isotope exchange, it is important to understand the disk dynamics necessary to account for the preservation of planet-scale and planetesimal-scale oxygen isotope heterogeneity in an environment where large-scale isotopic exchange processes are taking place.

**Nucleosynthetic isotope heterogeneity in planetary materials**

Much recent work on meteorites has focussed on observations of mass-independent, nucleosynthetic isotopic anomalies that show an isotopic division, or dichotomy, among planetary materials (Warren 2011; Dauphas and Schauble 2016; Kruijer et al. 2020; Bermingham et al. 2020; Kleine et al. 2020). Nucleosynthetic isotopic variations reflect the fact that the products of stellar nucleosynthesis were not fully homogenised in the protoplanetary disk. Isotopic diagrams that illustrate these effects (such as Fig. 7) are typically shown in epsilon units, ε, which reference the ratio of the isotope in question to another isotope of the same element, and to a standard, in parts per 10,000. As an example, $\varepsilon^{50}Ti = [(^{50}Ti/^{47}Ti)_{sample} / (^{50}Ti/^{47}Ti)_{standard} -1] \times 10,000$. One ε unit therefore represents a 0.01% deviation in the isotopic ratio of the sample relative to the standard.

Two distinct isotopic reservoirs have been defined, on the basis of isotope ratios for multiple elements. These are named the carbonaceous chondrite (CC) and non-carbonaceous chondrite (NC) reservoirs, since they are largely defined by carbonaceous vs other chondrite groups (Fig. 7). However, the division is not limited to chondrites alone: although most achondrites fall in the NC group, some ungrouped achondrites lie in the CC group (Sanborn et al. 2013 2019). In Figure 7, CC-group achondrite data points are for individual meteorites whereas the NC-group data points are averages for different large achondrite groups including HEDs, aubrites, angrites and ureilites. The two isotopic groups are clearly distinguished on plots of $\varepsilon^{50}Ti$ vs $\varepsilon^{54}Cr$ (Fig. 7a): $^{50}Ti$ and $^{54}Cr$ are neutron-rich isotopes thought to be produced in similar astrophysical environments, but the site and method of their production are still somewhat uncertain (Qin and Carlson 2016). The two isotopic groups are also commonly shown on plots of $\Delta^{17}O$ vs $\varepsilon^{54}Cr$ (e.g. Scott et al. 2018; Kruijer et al. 2020), although oxygen isotope variations are not nucleosynthetic in origin (see above).

Molybdenum isotopes are particularly useful for defining the two nucleosynthetic isotope reservoirs because Mo is measurable in a wider range of materials, including iron-rich meteorites (Fig. 7b). On a plot of $\varepsilon^{95}$Mo vs $\varepsilon^{94}$Mo (Fig. 7b), CC-group and NC-group meteorites define two parallel lines with similar slopes but different intercepts. The separate lines are s-process mixing lines, and the offset between the lines reflects a uniform r-process excess in the CC reservoir (Kruijer et al. 2020). An important observation is that there are iron meteorite and pallasite members of both the CC and NC isotopic groups (Kruijer et al. 2017a; Budde et al. 2019: Fig. 7b). The CC-group pallasites are the rare Eagle Station and Milton groups which represent less than 10 meteorites out of a total of 170 known pallasites. The CC-group iron meteorites are also some of the less abundant groups (see Fig. 4): the relevant groups are represented in total by <100 meteorites out of 1375 known iron meteorites. Given that carbonaceous chondrites are themselves a small proportion of all chondrites (5%, see above), our current sampling of extraterrestrial material is strongly weighted to the NC isotopic group. To date, we are only able to place two planets on the nucleosynthetic isotope diagrams, Mars and the Earth / Moon system (Earth and the Moon are essentially isotopically indistinguishable). Both Mars and Earth belong to the NC group (Fig. 7), consistent with their formation predominantly from material similar to enstatite and ordinary chondrites, and/or associated differentiated planetesimals. Enstatite chondrites are a match to the Earth for multiple isotopic systems, which leads to the suggestion that Earth and enstatite chondrites formed from the same isotopic reservoir, even though they are chemically distinct (Javoy et al. 2010; Dauphas and Schauble 2016, Dauphas 2017).

The prevailing explanation for the two isotopic reservoirs is that the protoplanetary disk was divided into two regions because growth of Jupiter created a gap in the disk, preventing mixing of inner disk (NC) and outer disk (CC) materials (Kruijer et al. 2017a, 2020; Desch et al. 2018; Bermingham 2020; Kleine et al. 2020). This model leads to a common assumption that the CC isotopic reservoir consists of outer disk material that has higher volatile element contents than those of the OC reservoir, and an inference that the two reservoirs are "wet" (CC) and "dry" (NC), forming outside and inside the water snow line. However, in detail the division is not so clear. As discussed above, there is ample evidence that ordinary chondrites and NC-group achondrites accreted water ice, and NC-group chondrites as well as ureilites accreted carbon, including organic material. Among the iron meteorites, CC-group irons and NC-group irons have variable and overlapping Ge contents (Fig. 4). Moderately volatile elements such as Mn, Na and K are depleted in most groups of carbonaceous chondrites relative to solar abundances, and to a greater extent than in ordinary chondrites. The enstatite chondrites also have relatively high abundances of moderately volatile elements, even though they are commonly assumed to have accreted close to the Sun. Hence, the inner Solar System was not completely devoid of volatile elements during planetesimal accretion and volatility trends are not systematic.

Recognition of the nucleosynthetic isotope dichotomy has sparked much discussion of disk dynamics that are relevant to observations of other young planetary systems undergoing planet formation. There is also debate about the origin of nucleosynthetic isotope anomalies, for example whether they represent addition or subtraction of presolar carrier components, such as supernova dust, to the inner vs outer Solar System, and when such processes might have taken place in relation to the timing of planet formation (e.g. Qin and Carlson 2016; Nagashima et al. 2018; Ek et al. 2020; Lichtenberg et al. 2021; Hopp et al. 2022). There is a need to reconcile such models with variation in oxygen isotope compositions, bulk chemical compositions, and the relationship of planetesimal or planet accretion zones to the snow line. If CC materials formed beyond the orbit of Jupiter, later perturbations of the system must account for the current presence of CC-type asteroids (both chondritic and differentiated) in the main asteroid belt. Also, disk models must account for the fact that physical processes, such as the heating events that formed chondrules, and differentiation of early-formed planetesimals, took place in both regions of the disk.

# TIMELINE OF EVENTS INFERRED FROM METEORITES

Meteorites and returned samples allow us to measure the chronology (timing) of the formation and evolution of Solar System bodies with high accuracy and precision. Two types of chronometers are used, both based on radiometric methods. Firstly, absolute chronometers provide a direct measurement of the date of a particular process, using radioactive decay schemes with long (billions of years) half-lives. For example, using the decay of uranium to lead, a precise formation age for calcium-, aluminum-rich inclusions (CAIs) has been determined, $4567.30 \pm 0.16$ Ma (Connelly et al. 2012). This age defines the timing of formation of the first solid particles within the protoplanetary disk, and it is taken to be time $t = 0$, as a reference for subsequent events. The second dating method uses relative chronometers, which are based on decay of short-lived radioisotopes that were present in the early Solar System. Since these isotopes have now decayed completely, it is not possible to use a conventional chronometer approach. Instead, they must be referenced to a measurement that anchors a precise absolute age to a known abundance of the radioactive isotope, relative to a stable isotope of the same element. As an example, decay of $^{26}$Al to $^{26}$Mg has a half-life of 0.7 m.y.. The ratio $^{26}$Al/$^{27}$Al, determined from CAIs, is $5.25 \times 10^{-5}$, and this is taken to be the reference ratio at $t = 0$ (Kita et al. 2013). A chondrule might have a $^{26}$Al/$^{27}$Al ratio of around $1 \times 10^{-5}$, indicating that it formed 2 million years after $t = 0$. The absolute age of the chondrule has not been measured, but the relative chronology is established. A second important short-lived chronometer, decay of $^{182}$Hf to $^{182}$W (half-life 8.9 m.y.) is used to establish timing of core formation (Kleine et al. 2009). In a molten planetesimal or planet, Hf, a lithophile (rock-seeking) element, partitions into the mantle and W, a siderophile (metal-seeking) element, partitions into the core. If core formation is completed before $^{182}$Hf has completely decayed, an excess of $^{182}$W will be observed in mantle materials. Use of relative chronometers relies on the assumption that the radioactive isotope is distributed homogeneously throughout the disk, and such assumptions are questioned repeatedly (e.g. Connelly et al. 2017; Nagashima et al. 2018; Bollard et al. 2019).

Detailed dating of planetary processes establishes the timeline of events in the protoplanetary disk. The older view that chondrites are the oldest materials, and that differentiated planetesimals formed later, has been overturned. There is now abundant evidence for early core formation, and it is clear from achondrites and iron meteorites that accretion and melting of many differentiated parent bodies took place within two million years after CAI formation, and that most iron meteorites accreted in the first million years (Kleine et al. 2009; Desch et al. 2018; Bermingham et al. 2020; Kruijer et al. 2020; Lichtenberg et al. 2021). This is consistent with early-formed planetesimals having a higher abundance of $^{26}$Al, an important heat source that led to extensive melting. Planetesimals that accreted later (e.g. 2-4 m.y. after CAIs) did not reach such high peak temperatures: these are the parent bodies of chondritic meteorites in which primitive disk materials are preserved. This model opens questions such as whether the materials that accreted to form differentiated planetesimals were chondritic, i.e. whether they included chondrules and matrix, and whether the chondrule formation period extended to a time prior to formation of differentiated planetesimals. (Some chondrule-formation models, such as formation in the bow shock of a planetesimal migrating through the disk, require planetesimals to have formed before chondrules: Morris and Boley 2018.) Most $^{26}$Al chondrule ages are around 1-4 m.y. after CAI formation (Kita et al. 2012; Nagashima et al. 2018), but Pb-Pb ages extend chondrule formation to older dates overlapping with CAI formation (Connelly et al. 2017). A further perennial question is how CAIs were stored for 1-2 m.y. until they mixed with other chondrite components to form chondritic planetesimals, and how they were distributed throughout the disk from their presumed initial locations close to the Sun (Desch et al. 2018; Jongejan et al. 2023).

Heating of chondritic planetesimals as a result of $^{26}$Al decay began soon after accretion. Aqueous alteration resulted in carbonate formation, at times of around 3-5 m.y. after CAI formation (e.g. Fujiya et al. 2012). Carbonate ages from asteroid Ryugu are significantly younger, 1.8 m.y. after CAI formation, indicating very early accretion and heating of the initial Ryugu parent body (McCain et al. 2023). More extensive heating, and slow cooling from the peak temperatures of metamorphism, such as in ordinary chondrites, continued for tens of millions of years (e.g. Blackburn et al. 2017). On differentiated planetesimals, achondrites such as angrites and eucrites crystallised within a few million years of CAI formation (e.g. Hublet et al. 2017). Since this early period of geologic activity, essentially for the last 4.5 billion years, the dominant geologic process on asteroid-sized bodies has been impacts. The record of impacts includes brecciation (i.e. solid-state fragmentation), heating to varying degrees above and beyond the point of melting, and high-pressure shock effects.

Accretion of the terrestrial planets occurred very early, likely within the first 10 million years after CAI formation (e.g. Kruijer et al. 2017b; Halliday and Canup 2023), and the Moon-forming giant impact could have taken place as early as 60 m.y. after CAI formation (Barboni et al. 2017; Thiemens et al. 2019). The oldest mineral grains, zircons, from the Earth and Moon record ages of 4.38 Ga and 4.4 Ga, respectively (Harrison 2009; Nemchin et al. 2009), constraining the timing of crust formation.

## METEORITES AND PLANET FORMATION: OVERVIEW

The above discussion illustrates how meteorites contribute significantly to our fundamental understanding of formation of planetary systems. They enable us to address the chemical evolution of disk materials, the chemical and isotopic heterogeneity of planets and planetesimals, and the timing of accretion and early geologic processes on rocky bodies. While fundamental major differences among the planets such as the relative sizes of planetary cores have been understood for a long time, it is only through laboratory studies of meteorites and returned samples that the details of mantle geochemistry can be surveyed.

Timing and location are fundamental to establishing the bulk chemical and isotopic properties of individual planets. Factors related to location of the planet-forming region include abundances of moderately volatile elements (such as Mn, Na, S, Ge) and volatile compounds such as ices ($H_2O$, $CO/CO_2$) and organic material, the availability of which would be expected to vary initially as a function of heliocentric distance, hence temperature, in the protoplanetary disk. However, any systematic, volatility-controlled chemical gradation in the disk is overprinted by planet migration, and the consequent perturbation of the distribution of small bodies (Walsh et al. 2011; Morbidelli et al. 2015; Nesvorný 2018). Identifying the feeder zones for the terrestrial planets is therefore a complex problem, as exemplified by continuing discussion of the chemical and isotopic nature of meteorite components that could have contributed to the bulk composition of the Earth and Mars (e.g. Dauphas and Schauble 2016; Budde et al. 2019; Mezger et al. 2020; Halliday and Canup 2023). Factors related to timing include the changing nature of available material as the snow line evolves, and the timing of planet formation in relation to formation of a Jupiter-related gap in the disk. Also, early-formed planetesimals are more likely to have differentiated, and could potentially have been degassed, prior to accretion to a planet (Dhaliwal et al. 2018), while later-formed chondritic planetesimals are more likely to preserve low-temperature components. Giant impacts between planet-sized bodies, such as the Moon-forming impact (Asphaug 2014; Lock et al. 2018), and impacts that could have removed a large part of Mercury's mantle (Asphaug and Reufer 2014; Chau et al. 2018), also contribute to more randomised planetary chemistry. This is particularly the case in the early stages of planet formation when major collisions are more likely. Therefore, interpretations of planet compositions, including exoplanets, from remote observations, need to bear in mind that the current composition could have been modified substantially from material that originally accreted.

To sum up, it is serendipitous that small bodies are preserved as asteroids in the inner regions of our Solar System. The asteroid population preserves protoplanetary disk material with a variety of compositions, as well as a detailed chronology, and delivers the record of it to Earth in the form of meteorites. If we were to attempt to understand the Solar System without the benefit of insights from meteorites, we would be ignorant of much of our current understanding, and we would likely have a very different perspective of the formation and evolution of the terrestrial planets. Even though we are only able to build this detailed picture for a single planetary system, the lessons learned can be applied to interpretations of other planetary systems, and the nature of exoplanets that they contain.

**Acknowledgements:** This work was partially supported by STFC grant ST/V000675/1.

# FIGURE CAPTIONS

Figure 1. Chondrite meteorites. a) CV3 carbonaceous chondrite, NWA 3118, 4 cm across. Circular objects are chondrules, white irregular objects are CAIs, dark material between chondrules and CAIs is the fine-grained matrix. b) Aba Panu L3 ordinary chondrite, 5 cm across. Circular objects are chondrules. Highly reflective material, white on the upper surface, is metal and sulfide which commonly surrounds chondrules. Striations on the upper surface are saw marks. c) False-colour combined X-ray map of L3 ordinary chondrite, OUT 18016: Mg red, Ca green, Al blue. Chondrules (circular objects) show different textures including barred olivine (large chondrule, lower left) and porphyritic (e.g. large chondrule centre right). Chondrule mesostasis is blue. Metal and sulfide are black; matrix is dark, interstitial to chondrules. X-ray maps courtesy of Jane MacArthur, Lost Meteorites of Antarctica project, The University of Manchester. d) False-colour combined X-ray map of CM2 carbonaceous chondrite, Murchison: Mg red, Ca green, Al blue. Chondrules appear red, CAIs are blue, carbonates are green. X-ray maps courtesy of Catherine Harrison, Natural History Museum / The University of Manchester.

Figure 2. Bulk chondrite carbon and water contents for ordinary and carbonaceous chondrites and for asteroid Ryugu which is comparable to CI1 chondrites. Data are from Alexander (2012), Vacher (2020), and a compilation of literature data in Piani et al. (2021), with additional data for Winchcombe (CM2) from Bates et al. (2023) and Verchovsky et al. (2023). Data for Ryugu and one data point for Ivuna (CI1) are from Yokoyama et al. (2023).

Figure 3. Differentiated and planetary meteorites. a) Eucrite NWA 11245, consisting predominantly of white grains of plagioclase feldspar and grey grains of pyroxene. 5 cm across. b) Eucrite, Northwest Africa (NWA) 1109. Plane-polarised transmitted light, field of view is 5 mm. NWA 1109 is brecciated, and is composed of comminuted fragments of individual mineral grains of plagioclase and pyroxene, as well as rock fragments (clasts). The larger coarse-grained rock clast on the right of the image consists predominantly of plagioclase feldspar (white) and pyroxene (brown) c) Angrite, Sahara (SAH) 99555. Plane-polarised transmitted light, field of view is 4 mm. Elongate white grains are feldspar, and pink / brown grains are pyroxene. d) Polished and etched surface of the iron meteorite, Gibeon, showing the Widmanstätten pattern and rounded inclusions of iron sulfide. Photo courtesy of Institute of Meteoritics, University of New Mexico. e) Lunar gabbro, NWA 8127. Cross-polarised light, field of view 1.25 mm. Igneous texture includes pyroxene (coloured grains with coloured lineations showing exsolution), olivine (rounded, coloured grains), oxides (black, angular) and maskelynite (shock-transformed plagioclase feldspar: black / grey). Fracturing in olivine and pyroxene results from shock. f) Martian meteorite, Tissint, an olivine-phyric shergottite (basalt). Plane polarised light, field of view 2.5 mm. Large grains of olivine are surrounded by a fine-grained groundmass consisting of pyroxene (grey), maskelynite (white) and oxides (black / opaque). Olivine and pyroxene are highly fractured from shock. The olivine grain in the upper right has two circular melt inclusions. The linear black feature in the upper left is an impact-melt vein.

Figure 4. Bulk compositions of iron meteorites. (Iron metal composition only for non-magmatic irons.) Magmatic irons are solid symbols; non-magmatic irons are open symbols. The CC isotopic group is in shades of blue; NC isotopic group is in shades of red and orange. Data sources are those listed in Table 1 of Goldstein et al. (2009).

Figure 5. Mn vs. Fe atoms per 4-oxygen formula unit (afu) for olivine, and per 6-oxygen formula unit for pyroxene, in basaltic igneous rocks from various bodies: HED achondrites (presumed from asteroid 4 Vesta), angrites, Mars, Earth, Moon. Slopes of lines are taken from Papike et al. (2003, 2017).

Figure 6. Oxygen isotope compositions of meteorites and their components. Reference lines are: TF, terrestrial fractionation line (slope = 0.52 on $\delta^{17}O$ vs $\delta^{18}O$, slope = 0 on $\Delta^{17}O$ vs $\delta^{18}O$); CCAM, carbonaceous chondrite anhydrous mineral line (slope = 0.94 on $\delta^{17}O$ vs $\delta^{18}O$); PCM, primitive chondrule mixing line (slope = 1 on $\delta^{17}O$ vs $\delta^{18}O$). a) Oxygen isotope compositions for bulk chondrites and asteroid Ryugu. Data points for carbonaceous chondrites are individual analyses; data points for the H, L and LL groups of ordinary chondrites, and Rumuruti (R) group chondrites, are means for individual groups with 2σ standard deviation indicated. b) Range of oxygen isotope compositions in chondrite components and the Sun. Solar value is from McKeegan et al. (2011). Open ellipses show fields for bulk compositions of chondrules in ordinary chondrites (red) and selected groups of carbonaceous chondrites that have not undergone extensive aqueous alteration (blue). These fields are similar to fields for bulk chondrites in (a): note difference in axis scales. The orange line shows the array of oxygen isotope compositions from individual in situ measurements of chondrule mineral grains, which includes OC and CC chondrules. The green line shows the array of oxygen isotope compositions from individual in situ measurements of grains in CAIs. The arrow points towards a "heavy", $^{16}$O-poor oxygen isotope component on the slope-1 line. c) Bulk meteorite oxygen isotope compositions for selected groups of achondrites, and martian meteorites. Data points are individual analyses apart from lunar basalts which is an average from Hallis et al. (2010). Most achondrite groups and martian meteorites lie on distinct linear arrays with different values of $\Delta^{17}$O. In contrast, ureilites show a wide scatter. d) Achondrite, lunar and martian meteorite data from (c), excluding ureilites. Data are from: Greenwood et al. (2017) and a compilation therein, Clayton and Mayeda (1999), Franchi et al. (1999), Hallis et al. (2010), Bischoff et al. (2011), McKeegan et al. (2011), Greenwood et al. (2015), Greenwood et al. (2023).

Figure 7. Nucleosynthetic isotope variations in meteorites: carbonaceous chondrite (CC) isotopic group in blue, non-carbonaceous (NC) isotopic group in red. Chondrite groups are filled circles, achondrites squares, pallasites diamonds, iron meteorites triangles, planets star symbols. a) $\varepsilon^{50}$Ti vs $\varepsilon^{54}$Cr. Data are from a compilation by Rüfenacht et al. (2023), also Trinquier (2009) and Sanborn et al. (2013, 2019). b) $\varepsilon^{95}$Mo vs $\varepsilon^{94}$Mo. Data are from a compilation by Budde et al. (2019), also Burkhardt et al. (2011).

**FIGURES**

**Figure 1**

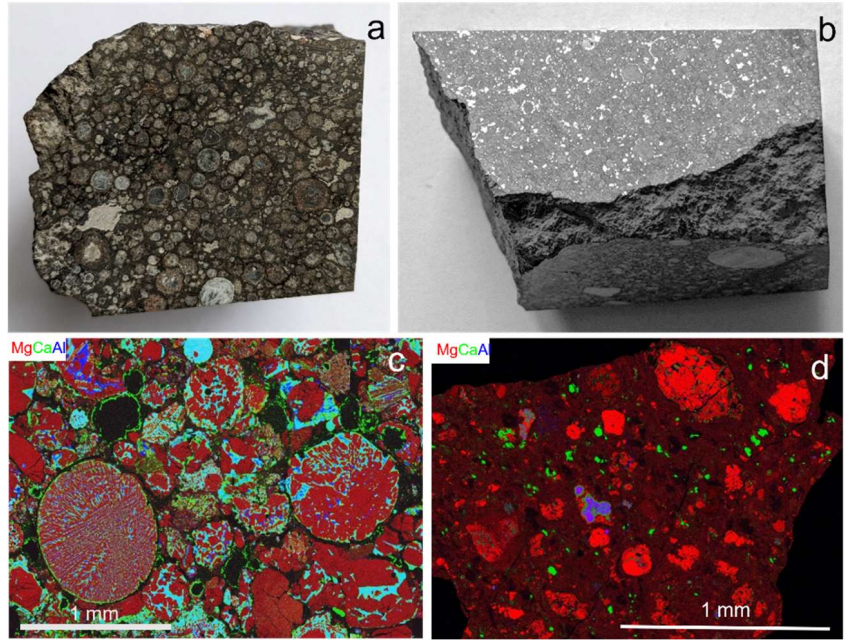

**Figure 2**

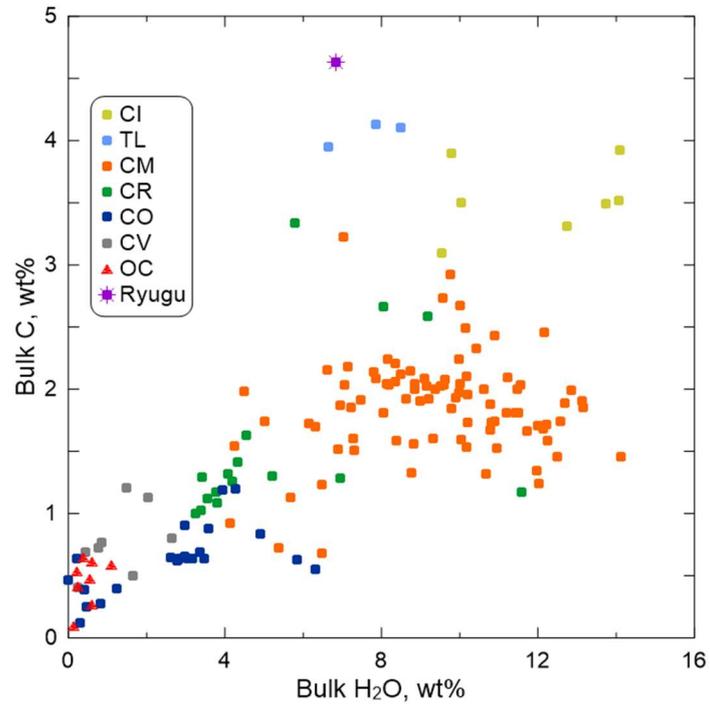

**Figure 3**

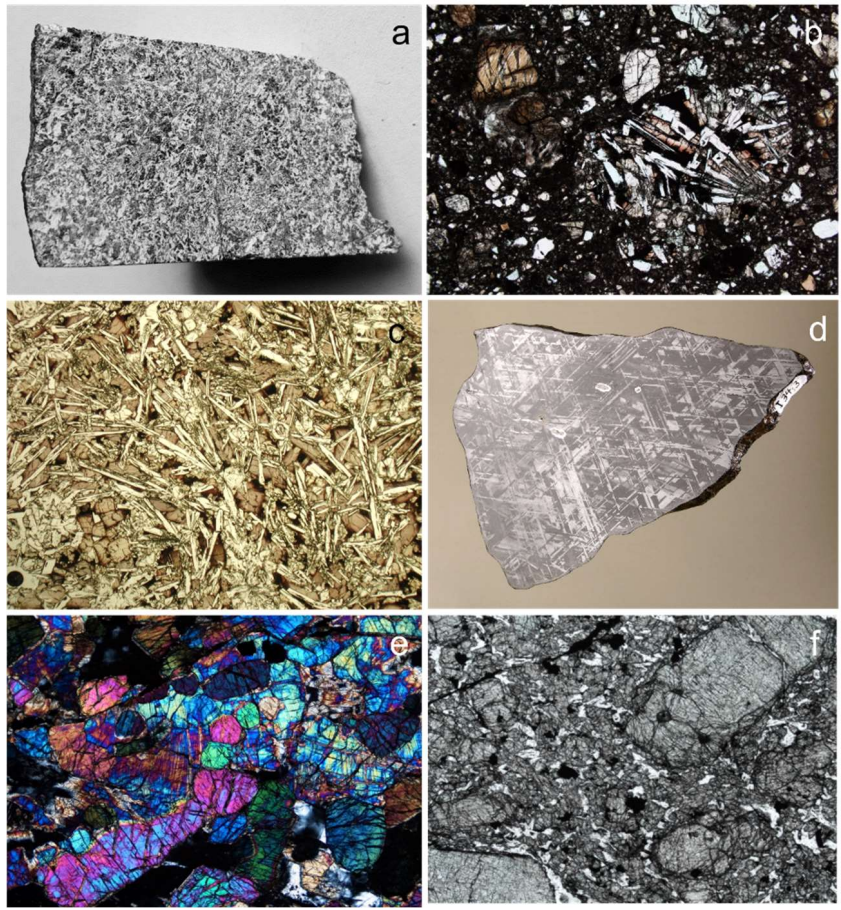

**Figure 4**

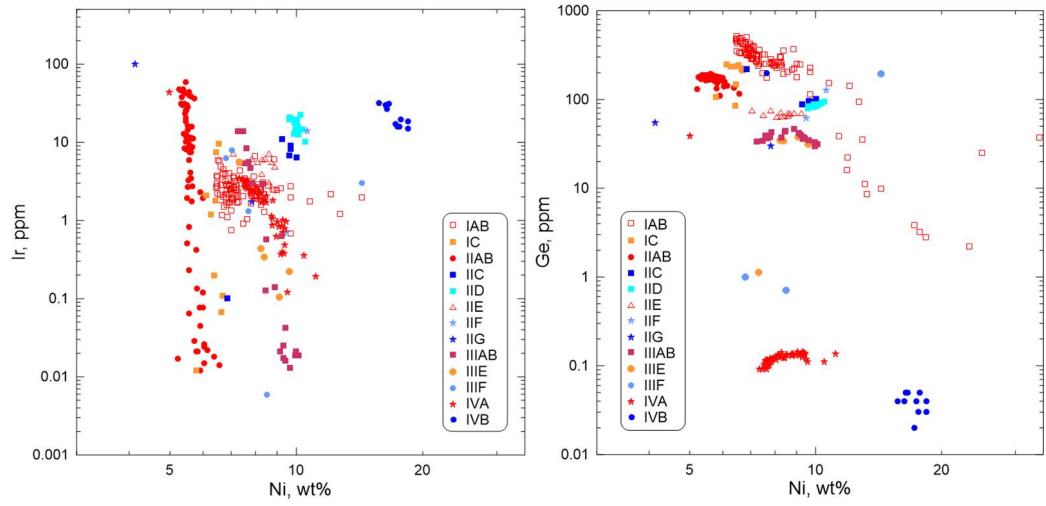

**Figure 5**

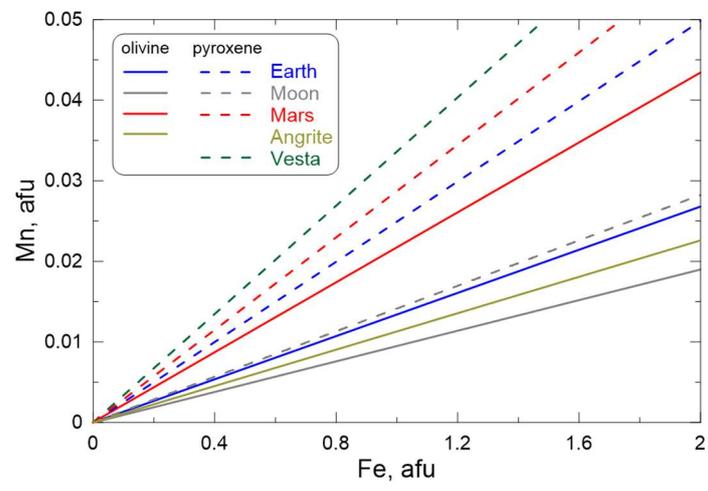

**Figure 6**

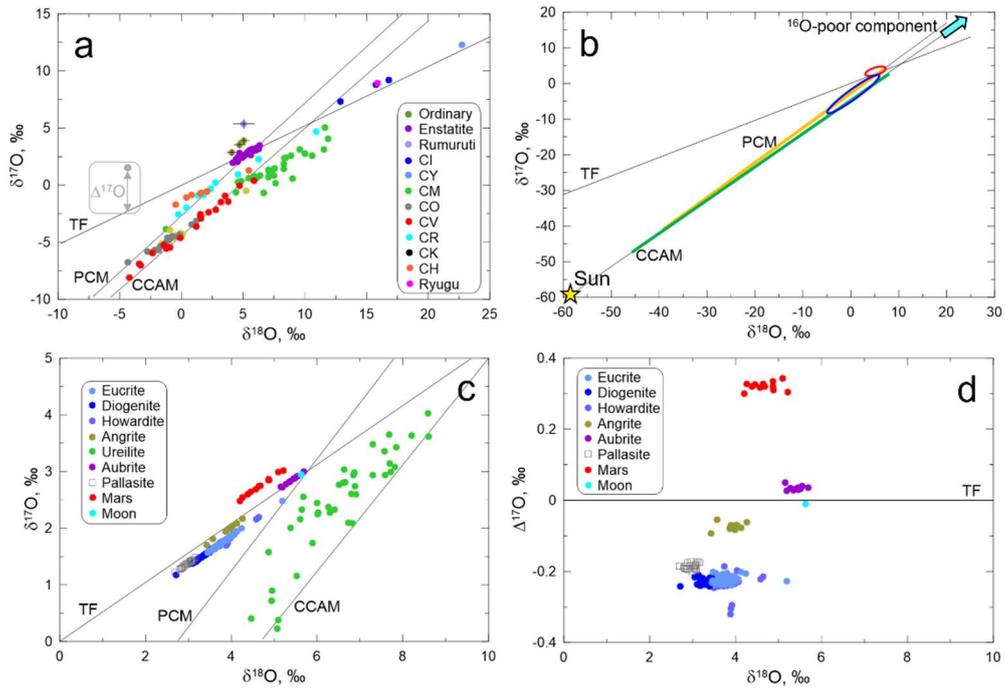

**Figure 7**

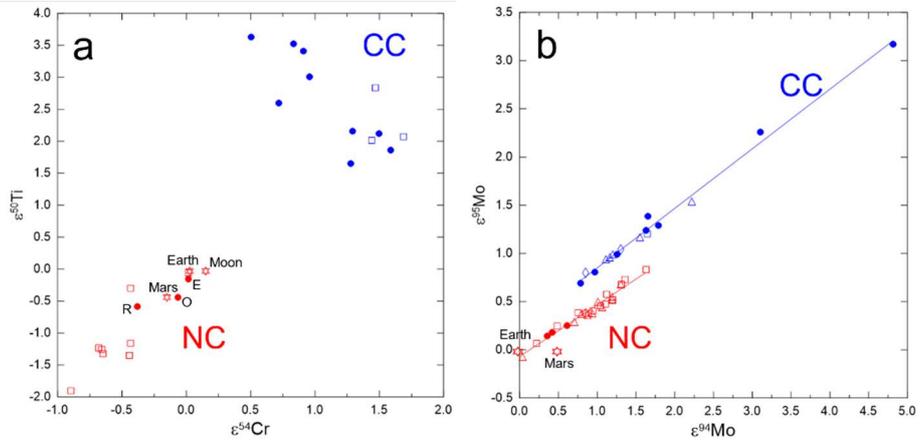